\documentclass[usenatbib]{mn2e}
\usepackage{graphicx}
\usepackage{lscape}

\begin{document}

\title{Star formation in Blue Compact Dwarf Galaxies : Mkn 104 and I Zw 97}

\author[S. Ramya et al.]{S. Ramya\thanks{E-mail : ramya@iiap.res.in (SR)}, D. K. Sahu\thanks{E-mail : dks@iiap.res.in (DKS)}, T. P. Prabhu\thanks{E-mail : tpp@iiap.res.in (TPP)} \\
Indian Institute of Astrophysics, Koramanagala, Bangalore-34, India.}

\date{Received / Accepted}

\pagerange{\pageref{firstpage}--\pageref{lastpage}} \pubyear{2006}

\maketitle

\begin{abstract}
Two blue compact dwarf galaxies Mkn 104 and I Zw 97 are studied photometrically and spectroscopically. Mkn 104 is found to contain three distinct bright star forming regions, whereas I Zw 97 is found to contain three bright and two faint star forming regions. Medium resolution spectra of three bright H {\sc ii} regions in the two galaxies were obtained. Estimation of oxygen abundance in these regions yields a value equal to log(O/H)+12 = 8.5 ($Z=Z_\odot/2.7$). Star formation rates in these star forming regions are estimated. The highest star formation rate for I Zw 97 is found to be 0.04 $M_\odot$yr$^{-1}$ and for Mkn 104, it is 0.02 $M_\odot$yr$^{-1}$. I Zw 97 is realized to be a cometary blue compact dwarf galaxy undergoing a strong burst of star formation. A $U-B$ vs $V-I$ colour-colour mixed population model is created using the Starburst99 evolutionary model curves. The spectrum of the bright star forming knot of I Zw 97 does not show any strong signature of an underlying relatively older stellar population, but the $U-B$ vs $V-I$ two colour diagram indicates a strong contribution of a $\sim500$ Myr population. A spectrum of the central region of Mkn 104 gives a hint about the underlying old stellar population. The age of this underlying population using the $U-B$ vs $V-I$ two colour diagram is estimated to be $\sim500$ Myr. Surface brightness profiles and colour profiles for these galaxies are presented. The surface brightness profile of both the galaxies can be represented well by a two-component S\'ersic profile consisting of a near exponential distribution and a Gaussian nuclear starburst. To conclude, both of these galaxies are not young systems; instead they are undergoing episodic star formation superposed on a faint older component. I Zw 97 is a cometary blue compact dwarf galaxy where the underlying low surface brightness (LSB) galaxy is a dwarf irregular observed during a major stochastic enhancement of its otherwise moderate star formation activity, a phenomenon widely accepted as self propagating stochastic star formation. Both these galaxies are very similar in their stellar content, showing an older 4 Gyr population, an intermediate 500 Myr population and the current burst of star formation of age 5-13 Myr.
\end{abstract}


\section{\bf INTRODUCTION :} 

Three decades of study of Blue Compact Dwarf (BCD) galaxies has led to a preliminary understanding of their content and evolution. Some of the key results have been summarized by \cite{sn99}: \\

\begin{enumerate}

\item{BCDs are dwarfs currently undergoing a strong burst of star formation (SF) whose optical light is dominated by emission from hot young stars.}
\item{Since star forming regions in BCDs are unusually bright, it becomes difficult to assign morphological types to the underlying hosts. But currently there has been substantial evidence that the underlying host is a dwarf irregular galaxy whose radial profiles are exponential at least at large radii (\citealt{mar99}, \citealt{meu00}).}
\item{Spectra of the BCDs are dominated by nebular emission superimposed on the featureless continua of O and B stars. The emission-line dominated spectra permit an accurate determination of elemental abundances of the emitting gas.}
\item{It has also been well established that BCDs are not very young systems forming stars for the first time; rather they all possess an older underlying population of stars which confirms that BCDs began forming stars long ago (\citealt{pap96a}, \citealt{pap96b}, \citealt{tt97}).}
\item{BCDs are deficient in CO \citep{bar00}, but this does not necessarily imply a lack of H$_2$. It has also been suggested that CO and even H$_2$ might have been destroyed in the extreme, compact environments of BCDs. Clearly more observations are required in this regard.}

\end{enumerate}

There are still many questions which need special attention. There are a few speculations which need to be resolved:

\begin{enumerate}

\item{It is not yet clear what initially triggers the burst of star formation. Gravitational instabilities, random cloud collisions in the interstellar medium (ISM), tidal interactions, collisions with other galaxies, ISM stripping and cooling flow accretion in specific galaxies are some of the star formation triggers as reviewed by \cite{bro99}.}
\item{Once a burst has occurred, it can be sustained and regulated by the underlying stellar distribution through shock waves from winds and supernova explosions of massive stars (\citealt{meu76}, \citealt{hel99}) which was generalized by \cite{ger80} as self-propagating stochastic star formation (SPSSF).}
\item{The major issue concerns the evolution of a dwarf galaxy (\citealt{thu85}, \citealt{lf83}). \cite{sn99} have shown that all gas-rich dwarf galaxies do not go through one or more BCD like episodes to end as gas poor dwarf ellipticals. Rather, a special subsample of the dwarf galaxy population can host a starburst of the magnitude that would qualify it as a BCD. Milder star bursts observed in these epochs may not be effective in expelling the gas so as to change a galaxy's morphological classification. In fact, the morphology of a dwarf galaxy is largely set by its enveloping dark halo \citep{meu00}.}

\end{enumerate}

In this paper, we have taken up a task to study the star formation properties and star formation history of two very different BCDs: Mkn 104 and I Zw 97, using broad band $U$$B$$V$$R$$I$ photometry, narrow band H$\alpha$ imaging and medium resolution spectroscopy of the intense star forming regions. I Zw 97 resembles a cometary BCD. RC3 (third reference catalogue of bright galaxies,\citep{devau91}) reports a redshift of 0.008399 with Heliocentric radial velocity of $2518\pm72$ kms$^{-1}$ and the distance to this galaxy is calculated to be 34.5 Mpc (H$_0$ = 73 kms$^{-1}$Mpc$^{-1}$). The morphological type of the underlying faint low surface brightness galaxy has not been classified. This galaxy does not belong to any catalogued group and has no nearby neighbours.

Mkn 104 differs from I Zw 97 in its environment. It is classified to be a member of a group, UZC-CG 94 (\citealt{mer05}, \citealt{fk02}) consisting of three members. UGC 4906, a Sa type galaxy, is the nearest ($37'$ distance) neighbour to Mkn 104. The Heliocentric velocity of Mkn 104 is $2235\pm2$ kms$^{-1}$ (2004 SDSS Data Release 2; from NED) and the distance to this galaxy is estimated to be 30.62 Mpc (H$_0$ = 73 kms$^{-1}$Mpc$^{-1}$). It has been suggested that Mkn 104 contains double nuclei in the merging phase (\citealt{gim04}, \citealt{maz93}, \citealt{maz91}). However, \cite{maz93} have shown that both the nuclei exhibit H {\sc ii} region like spectra.

This paper is structured as follows. Section 2 of the paper deals with observations and data reduction. Section 3 and 4 deal with the results and star formation history of the two galaxies and finally Section 5 concludes the study.

\section{\bf OBSERVATIONS AND DATA REDUCTION :}

I Zw 97 and Mkn 104 were observed using the 2m Himalayan Chandra Telescope, of the Indian Astronomical Observatory, Hanle, India, which is remotely operated from the Center for Research and Education in Science and Technology (CREST), IIA, Bangalore. Images of these galaxies were taken through Bessell's broad band $U$$B$$V$$R$$I$ and narrow band H$\alpha$ (BW=100 \AA) filters, using the central 2K$\times$2K region of a 2K$\times$4K SITe CCD (plate scale $0\farcs296$ per pixel) installed in the HFOSC system (Himalaya Faint Object Spectrograph and camera, http://www.iiap.res.in/iao.html). Imaging observations were done under photometric sky conditions. The seeing at $B$ band ranged 1$\farcs$8 to 2$\farcs$2 (see Table \ref{t1}). Medium resolution spectra of a few of the bright star forming regions of the galaxies were obtained using a $11'\times1\farcs92$ slit (\#167l) with a grism (\#7) providing coverage of 3800-6700 \AA \ with a dispersion of 1.46 \AA/pixel. The spectral resolution was $\sim86$ kms$^{-1}$ ($\sim11$ \AA) FWHM at the central wavelength of 5340 \AA. The centres of the star forming regions were determined using images taken through H$\alpha$ filter, which is an excellent indicator of strong star formation. Spectra were then obtained by aligning the slit at the centre of these star forming regions.

Data reduction was done using the standard tasks available within IRAF\footnote{Image Reduction \& Analysis Facility Software distributed by National Optical Astronomy Observatories, which are operated by the Association of Universities for Research in Astronomy, Inc., under co-operative agreement with the National Science Foundation} which includes bias subtraction, flat fielding, cosmic rays removal, aligning of the frames and finally aperture photometry of the star forming regions. The observed frames were calibrated using the standard fields of \cite{lan92} observed on the same night. Spectroscopic reduction included bias subtraction, flat fielding, extraction of the one dimensional spectra, wavelength calibration using the ferrous argon arc lamp spectrum and flux calibration using spectrophotometric standards. Standard stars Feige 34 and Feige 67 were used to flux calibrate spectra of the star forming regions of Mkn 104 and standard star Hz 44 was used to flux calibrate I Zw 97 spectra. A log of observations is presented in Table \ref{t1}. Figures 1 and 2 show the $B$ band and H$\alpha$ narrow band images of I Zw 97 and Mkn 104, respectively. Figure 3 shows the spectra of a few star forming regions that are numbered in Figures 1 and 2.

\section{RESULTS :}

\subsection{H$\alpha$ photometry :}

 H$\alpha$ line images were obtained by subtracting the scaled $R$ band image from the H$\alpha$ image following the procedure suggested by \cite{wal90}. Figures 1(b) and 2(b) show the continuum subtracted H$\alpha$ images of the two galaxies. I Zw 97 shows two strong star forming regions numbered as knots k1 and k2 and three other faint star forming regions (knots k3, k4 and k5) spread over the face of the galaxy. The H$\alpha$ image of I Zw 97 is particularly interesting, as four of the five star forming regions are arranged in a cometary structure, with the brightest knot k1 situated at the south-western end. Knot k3 is near the centre of this galaxy and it is very faint, indicating very little star formation. The brightest knot, labeled here as k1, is at a distance of 1.6 kpc ($1\farcs0$ corresponds to 167.25 pc at a distance of 34.5 Mpc) from the central knot k3. Similar cometary Blue Compact Dwarf galaxies Mkn 59 and Mkn 71 were studied extensively by \cite{noe00}.

Mkn 104 has three spatially resolved knots marked k6, k7 and k8 in Figure 2. Knot k6 is near the centre of this galaxy; knots k7 and k8 are at a distance of 1.04 kpc and 2.6 kpc (at a distance of 30.62 Mpc, $1\farcs0$ corresponds to 148.5 pc), respectively from the central knot k6. The faint knot k8 at the northern edge of the galaxy is very well detected even in the $B$ band image (Figure 2(a)). A total of eight star forming regions were identified manually in these two galaxies. The H$\alpha$ flux for each of these regions were obtained within an aperture set for each knot as shown in the Figures 1(b) and 2(b). Aperture selection was based on the criteria that each pixel has a minimum of 3$\sigma$ counts above the background level, and it also has enough signal to give $U$$B$$V$$R$$I$ magnitudes which are accurate to at least 0.1 mag. H$\alpha$ absolute flux calibration was done using the standard stars from \cite{oke90}. H$\alpha$ equivalent width is calculated using the expression given by \cite{wal90} and is presented in Table \ref{t2}. The star formation rate (SFR) calculated from H$\alpha$ luminosity \citep{kenni98} is as high as 0.04 $M_\odot$yr$^{-1}$ for knot k1 of I Zw 97 and 0.02 $M_\odot$yr$^{-1}$ for knot k7 of Mkn 104.

\subsection{Broad band photometry :}

Broad band $U$$B$$V$$R$$I$ images clearly show the cometary extensions for I Zw 97 and three spatially resolved star forming regions of Mkn 104 (Figures 1(a) and 2(a)). Apertures drawn in the H$\alpha$ images centered around the star forming regions were used to obtain the broad band magnitudes and colours. For sky subtraction, sky and galaxy background were estimated in an annulus just beyond the apertures set for the star forming regions. Broad band magnitudes of the star forming regions were calibrated using standard stars from the list of \cite{lan92}. Photometric errors calculated for the $U$$B$$V$$R$$I$ magnitudes lie in the range 0.05-0.09 mag. Table \ref{t3} gives the magnitudes and colours of eight star forming regions of these galaxies.

\subsection{Surface brightness and colour profiles :}
\label{sbcp}

The surface brightness (SB) distribution of the underlying galaxy was obtained by fitting ellipses to the isophotes of the galaxies. This was done using the task {\it ellipse} available within {\it stsdas}. The bright H {\sc ii} regions in the galaxies were masked in an iterative way. To begin with, an initial guess of the center of the ellipse, ellipticity, position angle were provided. The prominent H {\sc ii} regions (except the central one) were masked manually by visual inspection and the ellipse fitting was carried out. The parameters of the best fit ellipse were used to generate an approximate H {\sc ii} region free model of the galaxy with the help of {\it bmodel} task. A residual image was obtained by subtracting the H {\sc ii} region free model image from the original galaxy. This residual image which consists of only bright H {\sc ii} regions, was run through {\it imreplace} task to replace all the other pixels with 0 except for the H {\sc ii} regions. This H {\sc ii} region image with zero background was subtracted from the original galaxy to get the underlying galaxy component.

This process was iterated until a smooth H {\sc ii} region free image of the underlying galaxy was obtained. The original $B$ band image and the H {\sc ii} region free model image are shown at top left and top right frames of Figure 4, respectively. This final H {\sc ii} region removed galaxy image was fed into GALFIT fitting algorithm of \cite{peng02}. In Figure 4, GALFIT's model fit image is shown in the bottom left panel, and the residual image (original image minus GALFIT's model fit image) is shown in the bottom right panel. The ellipse fitting task was run on the best fit image of GALFIT.  Both the galaxies I Zw 97 and Mkn 104 were best fit with a double S\'ersic profile whose parameters are shown in Table \ref{t4}. Further details are given in Section \ref{sbpcp}.

\subsection{Long slit spectroscopy :}
\label{lss}

Medium resolution spectra of bright star forming regions of knots k1+k2 of I Zw 97 and knots k6 and k7 of Mkn 104 are displayed in Figure 3. The two star forming regions k1 and k2 of I Zw 97, could not be spatially resolved and hence their composite spectrum was extracted and is labeled k1+k2. All  three spectra show strong emission lines of hydrogen Balmer series and of other elements like oxygen, nitrogen, sulphur and argon. All the Balmer lines are corrected for underlying stellar absorption. Correction for the underlying stellar absorption was made assuming an absorption equivalent width of 2 \AA \ \citep{mc85} following \cite{hh99} and \cite{pop00}. This correction is essential, as the internal reddening estimate and the abundance calculation strongly depend on the fluxes of these Balmer lines (especially H$\beta$ flux). The internal reddening (reddening within the galaxy) can be estimated using the H$\alpha$/H$\beta$ flux ratio as suggested by \cite{kong02}, provided the H$\alpha$/H$\beta$ flux ratio is greater than the theoretical flux ratio of 2.87. The H$\alpha$/H$\beta$ flux ratio for all the three regions yielded a value less than the theoretical ratio of 2.87 and hence our spectra are not corrected for internal reddening, but corrected for foreground Galactic extinction given by \cite{sch98}, as provided by NED. Oxygen abundances for all the three regions were calculated using the standard bright line methods as suggested in the combined method of \cite{kd02}. Flux ratios of various lines and the oxygen abundance estimated using these line ratios are listed in Table \ref{t5}. Mean log(O/H)+12 is 8.5 ($Z_\odot$/2.7 for solar oxygen abundance of 8.93) for the three regions. This value is subsolar, but not low enough to consider the current starburst as the first episode of star formation in these galaxies. The presence of an older population is evident from the presence of Balmer absorption lines superposed over the strong emission lines in the central knot spectrum (k6) of Mkn 104. This indicates that Mkn 104 might be hosting a substantial older stellar population underlying the current starburst. The [O {\sc iii}]/H$\beta > 1$ is obtained for all the three star forming regions implying harder ionizing radiation, a common feature found in other BCDs also \citep{hh99}.

\section{DISCUSSIONS :}

\subsection{Surface Brightness and colour profiles :} 
\label{sbpcp}

The H {\sc ii} region subtracted, smooth underlying galaxy image obtained through the iterative process (described in Section \ref{sbcp}), was fed into the GALFIT algorithm of \cite{peng02}. The surface brightness profiles of the two galaxies, I Zw 97 and Mkn 104 can be fit well using the S\'ersic profile. The S\'ersic profile is defined as :

{\ensuremath $\Sigma$(r) = $\Sigma_e$ $e^{-\kappa[(r/re)^{(1/n)}-1]}$}

where $r_e$ is the effective radius of the galaxy, $\Sigma_e$ is the surface brightness at $r_e$, $n$ is power-law index and $\kappa\sim2n - 0.331$. A S\'ersic profile transforms into a Gaussian when $n$ = 0.5, an exponential profile when $n$ = 1 and a de Vaucouleurs profile when $n$ = 4. Here, the general form of the S\'ersic profile was used to fit the surface brightness distribution of both the galaxies and it was found that two components are required for the fit.

The GALFIT fitting process was started with two S\'ersic profiles with power-law index $n$ = 0.5 and $n$ = 1 as an initial guess. The $n$ = 1 profile would represent the overall galaxy distribution if it is exponential and the $n$ = 0.5 profile would be able to fit the nuclear starburst found in both these galaxies. The best fit parameters obtained using GALFIT for both the galaxies are shown in Table \ref{t4}.

The surface brightness and colour profiles are plotted in Figure 5 for both the galaxies. In the plots, points correspond to surface brightness and colours of the galaxies with H {\sc ii} regions and thin lines correspond to the surface brightness and colours for the H {\sc ii} region subtracted smooth galaxy obtained using GALFIT. The surface brightness distribution of I Zw 97 can be modeled with an exponential profile except for a large bump at $\sim8\farcs0$ from the nucleus due to an intense starburst (Figure 5). This large bump is not seen in the modeled profile of the H {\sc ii} region masked galaxy obtained using GALFIT. The $B$ band surface brightness profile can be represented by a two component S\'ersic profile with the first component having power-law index $n\sim0.6$ (see Table \ref{t4}), a scale length of $\sim1$ kpc and surface brightness at $r_e$ of 20.78 mag.arcsec$^{-2}$ while the other component has power-law index $n\sim1.01$, a scale length of $\sim1.89$ kpc and surface brightness at $r_e$ of 22.64 mag.arcsec$^{-2}$. The centre of the first component is slightly offset from the centre of the second component by $\sim1\farcs3$ ($\sim220$ pc). This represents a starburst with a Gaussian distribution slightly offset from the central region of the galaxy. $I$ band surface brightness and colour profiles are affected by fringing due to which sky estimation becomes difficult and the outer isophotes are poorly detected.

The $B$ band surface brightness profile of Mkn 104 can also be represented by two component S\'ersic profile, one having a S\'ersic index of $n\sim1.72$ and another with S\'ersic index $n\sim0.93$ (the power-law index $n$ averaged over $B$$V$$R$$I$ bands is equal to $n\sim1.5$ for the one component and $n\sim0.65$ for the other component). The surface brightness at $r_e\sim0.45$ kpc is 20.20 mag.arcsec$^{-2}$ and at $r_e\sim1.04$ kpc is 21.42 mag.arcsec$^{-2}$ (Table \ref{t4}) for the $n\sim0.65$ and $n\sim1.5$ components in $B$ band, respectively. Both the components of Mkn 104 have S\'ersic index values higher than exponential and Gaussian values. This implies that light profile falls off slightly faster in Mkn 104 as compared to that in I Zw 97. The GALFIT results can be understood as large central at the centre of the galaxy, underlying which is a fainter continuum from a slightly older stellar population spread out over the face of the galaxy. A bump seen at radii of $5-6\farcs0$ corresponds to the H {\sc ii} region marked as knot k7 in Figure 2 and which is masked completely using the iterative method. The major axis diameter at 25.0 mag.arcsec$^{-2}$ in $B$ band is found to be equal to 38 arcsec, a value similar to the one reported in 2004 SDSS Data Release 2 (from NED). Here again the errors in the $I$ band profiles are confounded due to the fringing.

Figure 5 also shows colour profiles obtained by subtracting two surface brightness profiles. The colours, on average, remain constant until a semi-major axis length of about $5\farcs0$. Except for $V-R$ colour, $B-V$, $V-I$ and $R-I$ colours tend to become redder by 0.05-0.09 mag in the outer regions of both the galaxies. Photometric errors may be considerably large at outer isophotes and the observed variations may not be real. Colour profiles can be used as stellar age estimates when combined with a model such as Starburst99. Figure 6 shows a two colour diagram wherein each colour is calculated by subtracting two surface brightness profiles as obtained in Figure 5. Also plotted in these diagrams are Starburst99 \citep{leith99} colour-colour plots (thick lines in Figure 6) for metallicity ($Z$) 0.008 and mass range 0.1 $M_\odot$ - 100 $M_\odot$ undergoing an instantaneous burst of star formation and evolving from 0.01 Myr to 5 Gyr . If, in the plots, various colours ($B-V$, $V-R$, $V-I$ and $R-I$) represent the first and second dimension of the data, the third dimension for the plotted data points would be the radius or semi major axis length, while the third dimension for the thick line of Starburst99 would be the age. The concentration of points at a particular location in these diagrams are the central region points where the sampling of the isophotes is better than in the outer regions. A central concentration of points in $B-V$ vs $V-R$ plot is seen to coincide with an age of $\sim700$ Myr for I Zw 97 and $\sim550$ Myr for Mkn 104, which provides an estimate of the age of the underlying stellar population. The colours obtained from the surface brightness profiles are azimuthally averaged over an entire isophote and these colours may represent the galaxy background away from the current star forming regions.

\subsection{Star formation history :}

 To understand the star formation history of the two BCDs, we have used the colour-colour evolutionary population synthesis model of Starburst99 (\citealt{leith99}, \citealt{vl05}). The predicted colours evolving from 0.01 Myr to 5 Gyr were obtained using a proper choice of IMF and also the metallicity. We chose a Salpeter IMF with the lower and upper mass limits of 0.1 $M_\odot$ and 100 $M_\odot$. The metallicity ($Z$) of the galaxy is found to be 0.008 as measured from spectroscopy (Section \ref{lss}). The Padova 1994 stellar evolutionary models with AGB tracks were used to obtain the starburst colour tracks and SEDs in Starburst99 (dust-free model). The importance of the development of AGB stars at later ages (100 Myr - 1 Gyr) has been extensively reviewed by \cite{gir98}, \cite{gir02} and \cite{bc03}. It has been observed that the $U-B$ and $B-V$ colours are not much affected by the onset of the AGB for near-solar metallicity but the visual colours of the low-metallicity models are strongly affected by the development of the AGB as well as by all the uncertainties in the AGB evolution \citep{gir98}. The evolutionary tracks in Startburst99 created using these inputs are in very good agreement with that of the \cite{bc03} GALAXEV models. One such evolutionary track is plotted in Figure 7 for $U-B$ vs $V-I$ colours. $R$ band is not considered here because of uncertainties due to the H$\alpha$ emission present within $R$ band. The solid line track shown in Figure 7 runs from 0.01 Myr to 5 Gyr. Also plotted in the figure is the reddening vector for A$_V$ = 1 mag extinction. It is observed that the star forming knots of both the galaxies do not lie on the track exactly. The reddening vector clearly shows that the colours are reddened by 1 magnitude. The reason for this disagreement could be the uncertainties in the estimation of the galaxy background underlying these individual star forming regions which may lead to inaccurate flux estimates and hence colours. The second reason could be improper determination of the reddening values, and the third reason could be that there are multiple populations embedded in these regions. It has been well established now that BCDs are not very young systems undergoing their first episode of star formation (\citealt{pap96a}, \citealt{pap96b}, \citealt{tt97}, \citealt{sn99}, \citealt{noe00}), but rather possess an older, underlying population of stars. The reddening of 1 mag gives us a hint about the presence of an underlying old stellar population, although a small contribution due to dust within the star forming regions cannot be ruled out. This idea of underlying old stellar population may be incorporated in the mixed population colour-colour diagram to reveal further information about the star forming regions of these BCDs.

\subsubsection{\it Mixed population model for I Zw 97 :}

 The episodic star formation scenario was explored using the mixed population colour-colour model curve shown in Figure 7:

Consider a mixture of population ages ranging from 2 Myr to 9 Myr and a population ranging 200 Myr to 900 Myr with an underlying older stellar population of few Gyrs; then the combined colours of these three populations form a locus of points which are shown as smaller curves in Figure 7. Shorter curves in Figure 7 represent the locus of points ranging from 2 Myr + 200 Myr to 9 Myr + 900 Myr population mixed with a 4 Gyr old population. The fraction of young to older population can also be changed which gives 3 smaller curves displaced by small amounts. A burst strength parameter, $f$, is defined as the fraction of young population to old population by mass. The parameter $f_{2Myr}$ represents the fraction of young stars of age 2 Myr-9 Myr to older stars of age 4 Gyr. Similarly, parameter $f_{200Myr}$ is the fraction of 200 Myr-900 Myr old stars to 4 Gyr population. The shorter curves in Figure 7 represent loci of points corresponding to a mixed population of 2 Myr+200 Myr going to a 9 Myr+900 Myr with fraction $f_{2Myr}$ as 0.007 (dark blue curve), 0.01 (dark green curve) and 0.015 (magenta curve), keeping $f_{200Myr}$ constant at 0.05 (i.e., 5\% by mass), or in other words, the contribution of current burst is taken as 0.7\%, 1\% and 1.5\% by mass with respect to 4 Gyr old population. Knots k1, k2, k4 and k5 reside in these curves providing clues about their underlying stellar populations. A short dark red curve sits at a slightly redder $U-B$ colour. Knot k3, located at the centre of I Zw 97, is situated in this short dark red curve. This curve represents the locus of points having age range from 2 Myr+200 Myr to 9 Myr+900 Myr mixed with a 4 Gyr population, but the fraction of young to old, i.e $f_{2Myr}$ = 0.005 and fraction of 200 Myr to old, i.e $f_{200Myr}$ is 0.7. This implies that only 0.5\% by mass is in the current burst and a significant contribution ($\sim70$\%) is from a $\sim500$ Myr population. It can be clearly seen that all the knots of I Zw 97 are located on these curves. This substantiates the fact that this BCD has underlying older populations of age 4 Gyr + $\sim500$ Myr, and superposed on this is a much younger episode of star formation only 4.5-6 Myr. Knot k3 in the figure corresponds to the nucleus of I Zw 97 and is found to have a slightly older age $\sim5.7$ Myr for the current burst and knot k1 seen at the south-western end of the galaxy (Figure 1) shows a young 4.7 Myr population. Such strong extranuclear starburst regions are a signature of propagating star-forming activity along the main stellar body of these cometary BCDs, as also observed by \cite{noe00}.

The ages of the knots can be derived with better accuracy using the H$\alpha$ equivalent width versus age plot from a Starburst99 model for 0.008 ($Z$) metallicity. Figure 8 shows a plot of H$\alpha$ equivalent width versus age; it can be seen that the knots lie in the age range 6 Myr to 13 Myr (closed circles). However, the continuum around H$\alpha$ emission is contaminated by the old stellar population embedded in these regions. H$\alpha$ equivalent widths were corrected for underlying stellar populations of ages 500 Myr and 4 Gyr, as estimated above, resulting in younger ages for the knots (Figure 8; open circles). The ages of knots k1-k5 are in the range 5-6 Myr. The ages of star forming regions are presented in Table \ref{t2}. The estimated ages are similar to the ones obtained from the colour-colour mixed population model (Figure 7).

An underlying old stellar population aged 4 Gyr + 500 Myr is found in this galaxy, but the spectrum of the star forming knot k1+k2 shows no signatures of the underlying old component. Absorption lines like Mg I b absorption, Na D1 and D2 lines, CN band, characteristic of the Gyr old population are completely absent in the spectrum and, if present, have equivalent widths $< 1$ \AA. The reason for this could be the recent intense starburst at the locations of knot k1+k2. For a sample of BCDs, using an empirical population synthesis method with different continuum fluxes and absorption equivalent widths of the above mentioned lines, \cite{kong03} obtained the percentage contribution of newly born stars (or H {\sc ii} regions with $Age < 10^7$ yr), young stars ($10^7 < Age < 5\times10^8$ yr), intermediate age stars ($10^9 < Age < 5\times10^9$ yr) and old star ($Age > 10^{10}$ yr). According to this estimation, for I Zw 97 \cite{kong03} arrived at a conclusion that 60\% of the total light is due to stars of ages in the range $10^7 - 5\times10^8$ yr, implying population of young stars to be high. Intermediate age stars (1 Gyr- 9 Gyr) accounts for up to 20\%, and the remaining 20\% of the population involves newly born stars and very old stars within the galaxy. The contributions estimated by \cite{kong03} are for the light emission, whereas mass fractions are estimated here. The spectrum of the knot k1+k2 region does not show any absorption as mentioned above, but the two colour diagram hints at the presence of three distinct populations of ages $\sim5$ Myr, $\sim500$ Myr and 4 Gyr. Star formation rates calculated \citep{mou06} for the knots are as high as 0.04 $M_\odot$yr$^{-1}$. \cite{kenni84} measured masses and luminosities of a diverse sample of H {\sc ii} regions and found a reasonable empirical correlation between the two properties. \cite{kenni88} gave a linear conversion factor which is accurate at best to a factor of 3-4. Following \cite{kenni88}, the ionized gas mass of each of the knots are calculated and shown in Table \ref{t2}. The mass of the ionizing stars ($M_*$=10-100 $M_\odot$) in the whole galaxy is derived to be $7.05\times10^4$ $M_\odot$ and the mass of the ionized gas, M$_{\rm H II}$ is found to be $2.45\times10^6$ $M_\odot$. For comparison, \cite{men99} obtained a value of $7.2\times10^5$ $M_\odot$ as the mass of the ionizing stars, and $7.2\times10^6$ $M_\odot$ as the total H {\sc ii} mass, with a total star formation rate of 2.825 $M_\odot$yr$^{-1}$ derived for the wolf-rayet blue compact dwarf galaxy Mkn 1094. The emipirical relation involving $H$ and $K_s$ band magnitudes given in \citep{kir08}, \citep{vad07} is used to calculate the masses of the old stellar population. Total magnitudes as catalogued in 2MASS are used to obtain absolute $H$ and $K_s$ (M$_H$ and M$_K$) magnitudes. The mass-to-light ratio $\gamma_H$ and $\gamma_K$ values are taken as 1.0 and 0.8 respectively, as given in \citep{kir08} and \citep{vad07}. Using equation 8 of \citep{vad07} and standard equations in \citep{kir08}, $H$ band and $K_s$ band luminosities, total stellar mass, M$_{old}$ of I Zw 97 are calculated and M$_{old}$ is equal to $\sim2.1\times10^9$ M$_\odot$ units (refer Table \ref{t2}). This value is typical of many star forming dwarf galaxies.

\subsubsection{\it Star formation in Mkn 104 :}

Figure 7 also places the knots k6, k7 and k8 of Mkn 104 on the $U-B$ vs $V-I$ plot. It is observed that the star forming knot k7 is similar to the knots in I Zw 97, residing in the magenta curve which is a mixture of 2 Myr + 200 Myr and 9 Myr + 900 Myr populations mixed with a 4 Gyr population and at a fraction, $f_{2Myr}$=0.015, $f_{200Myr}$=0.05. This implies that knot k7 has an underlying older population of $\sim500$ Myr + 4 Gyr. Central knot k6 shows a larger $\sim500$ Myr population as it resides in the dark red curve where the central knot k3 of I Zw 97 is also placed, implying that the contribution from $\sim500$ Myr population is around 70\% by mass and of the younger population is around 0.5\% by mass. Knot k8 does not lie on the shorter curves, but is consistent with a similar older population considering various errors inherent to the estimation of broad band magnitudes, namely background and reddening correction. A strong contribution of underlying $\sim350-500$ Myr population is also seen in the spectrum of central knot k6, wherein all the hydrogen balmer lines are having an absorption component underneath the strong starburst emission (see Figure 3). Using the mixed population model, the ages of young components of the knots k6 and k7 are inferred to be 4 Myr and 6 Myr, respectively. Due to a stronger contribution of a $\sim500$ Myr population, we are at a loss to model the age of the young burst of knot k8 using the colour-colour diagram. Nevertheless, ages of the knots can be calculated with little uncertainty from H$\alpha$ equivalent widths and Mkn 104 knots k6-k8 are found to lie in the age range 7-13 Myr after correcting for an underlying older population of $\sim500$ Myr + 4 Gyr as shown in Figure 8. Here also, the central knot k6 is found to have a slightly older age of 13 Myr, and knot k8 which is a fainter knot situated at the northern end of the galaxy, is aged 7 Myr, hinting towards a SPSSF scenario. The star formation rate calculated is as high as 0.02 $M_\odot$yr$^{-1}$ for knots k6 and k7. The total mass of ionizing stars in the whole galaxy following \cite{kenni88} is found to be $1.04\times10^5$ $M_\odot$ and the ionized gas mass, M$_{\rm H II}$, is calculated to be $3.62\times10^6$ $M_\odot$ and total stellar mass is calculated to be $\sim1.43\times10^9$ M$_\odot$ units (refer Table \ref{t2}). Both I Zw 97 and Mkn 104 show a similar type of stellar content, wherein a very young burst with age $\sim5-13$ Myr, an intermediate population of age $500$ Myr and an old population of $\sim4$ Gyr are seen.

Further, \cite{izo06} obtained a value of 8.22 for the oxygen abundance and -1.23 for the log(N/O) ratio for Mkn 104. An increase in the N/O ratio signifies an enhancement of nitrogen by massive stars from the most recent starburst. \cite{izo06}, from a SDSS sample of $\sim400$ dwarf emission line galaxies, could not detect any galaxy having log(N/O) ratio $\le-1.6$ indicating that there were no young galaxy systems in their sample and that most galaxies had an intermediate log(N/O) ratio. With an intermediate value of -1.23 for log(N/O) ratio, Mkn 104 is not a young system, but rather a galaxy of age $>100-300$ Myr old as is required for the enrichment of nitrogen to such a value by intermediate mass stars \citep{izo06}. The two colour mixed population model created using Starburst99 arrives at the same conclusion.

Knots k6 and k7 of Mkn 104 have been considered to be double nuclei in a merging phase (\citealt{nor95}, \citealt{gim04}, \citealt{maz93}, \citealt{maz91}). However, Mkn 104 can be adequately modeled as a single galaxy having multiple starburst regions. \cite{maz93} have taken the spectra of the nuclei and confirmed that both the spectra are H {\sc ii} region like spectra. If such is the case, the reason of such a strong starburst needs to be understood. Mkn 104 is a member of a group, UZC-CG 94. This group has three members and the main member of the group is a Sa type galaxy, UGC 4906. This galaxy is $37'$ (or 330 kpc) away from Mkn 104. Presence of any interaction between the group members can only be confirmed through HI studies. Here we can only suggest that Mkn 104 is undergoing episodic star formation. A strong burst of star formation occurred $\sim500$ Myr ago and the galaxy is experiencing another milder episode of star formation in the present epoch.

\section{CONCLUSION :}

A study of broad band $U$$B$$V$$R$$I$ images, narrow band H$\alpha$ images and medium resolution spectroscopic observations of the BCDs Mkn 104 and I Zw 97 has been presented. Medium resolution spectra of the knots give an oxygen abundance estimate of $\sim8.5$ ($Z = Z_\odot/2.7$, (log(O/H)+12)$_\odot$=8.93) for both these galaxies. Surface brightness profiles in $B$$V$$R$$I$ bands show an exponential distribution of the underlying host and a Gaussian distribution is also seen centered around nuclear starburst regions for both these galaxies. For I Zw 97, the effective radius for the exponential distribution is $\sim1.89$ kpc in $B$ band, and the surface brightness is 22.64 mag.arcsec$^{-2}$ at this effective radius. The $B$ band surface brightness profile of Mrk 104 has a S\'ersic index of 1.5  and a surface brightness of 20.20 mag.arcsec$^{-2}$ at an effective radius of 0.45 kpc. The surface brightness profile of Mkn 104 falls slightly faster than that of I Zw 97. The centrally condensed system appears similar in both these galaxies, and with a near Gaussian shape and 1 kpc scale length it may represent the current starburst around the nucleus. Colour profiles were used to estimate the ages of underlying exponential distribution with the utilization of Starburst99 modeling, and the age of the underlying component was estimated to be $\sim700$ Myr for I Zw 97 and $\sim550$ Myr for Mkn 104. A relatively large effective radius for the elongated structure around star forming regions, and an underlying older (4 Gyr) population even after background subtraction, indicates that a bulge/bar like component may be underlying the star forming regions. This may be a common feature with BCDs, and could not be modeled here due to the bright star forming regions. \\

A mixed population colour-colour model created using the Starburst99 model curves is used to explain the broad band colours of the star forming knots of the two galaxies. I Zw 97 is modeled to contain three distinct populations with ages 5 Myr + 500 Myr + 4 Gyr. The age of the young population in these star forming regions is found to lie in the range 4.7-6 Myr. The central knot k3 is estimated to have a higher fraction ($\sim70$\%) of a $\sim500$ Myr old population and a small fraction ($\sim0.5$\%) of newly born stars with age $\sim6$ Myr. Knot k1, situated at south-west end of I Zw 97, has a younger population of age $\sim4.7$ Myr. This suggests self propagating stochastic star formation wherein the star formation first occurs at the nucleus and propagates through the stellar body of the galaxy. The knots of Mkn 104 show a strong underlying component of age $\sim500$ Myr and the younger bursts are in the age range 4-6 Myr as seen in the colour-colour mixed population plot (Figure 7). Both galaxies are undergoing a similar evolution wherein the oldest population in the two galaxies is $\sim4$ Gyr, a strong contribution of an intermediate population ($500$ Myr) is clearly seen, and newly born stars having age range $5-6$ Myr are also realized. Ages are calculated using H$\alpha$ equivalent widths after correcting for the estimated continuum from the underlying old and intermediate populations. Ages of the three star forming knots of Mkn 104 lie in the range 7-13 Myr. Ages of the knots of I Zw 97 calculated using H$\alpha$ equivalent width lie in the range 5-6 Myr. The SFR calculated using the H$\alpha$ luminosity is as high as 0.04 $M_\odot$yr$^{-1}$ and 0.02 $M_\odot$yr$^{-1}$ for the brightest star forming regions of I Zw 97 and Mkn 104 respectively. Both galaxies show indication of a significant population of age 300-500 Myr, which is enhanced in Mkn 104 possibly due to a group environment. I Zw 97 shows a higher rate of star formation at the current epoch, and indication of supernovae-induced star formation (I Zw 97 hosts the Type II supernova SN 2008bx, located close to knot k5, as discovered very recently by \cite{puc08}). It is possible that we are witnessing I Zw 97 at an earlier epoch of episodic star formation whereas Mkn 104 is in a later epoch. Alternatively, a stronger central component may sustain star formation longer in I Zw 97. 

\section{ACKNOWLEDGEMENTS :}
We thank the anonymous referee for careful and detailed comments which has led to the improvement of the manuscript greatly.
This work has made use of the NASA Astrophysics Data System and the NASA/IPAC Extragalactic Database (NED) which is operated by Jet Propulsion Laboratory, California Institute of Technology, under contract with the National Aeronautics and Space Administration. This research has made use of the NASA/ IPAC Infrared Science Archive, which is operated by the Jet Propulsion Laboratory, California Institute of Technology, under contract with the National Aeronautics and Space Administration. RS would like to thank University Grants Commission, GOI for their financial assistance. Authors wish to thank the support given by the staff of IAO and CREST, IIA.

\newpage

\begin{figure*}
\begin{minipage}{150mm}
\centering
\begin{tabular}{cc}
\includegraphics[width=8.5cm]{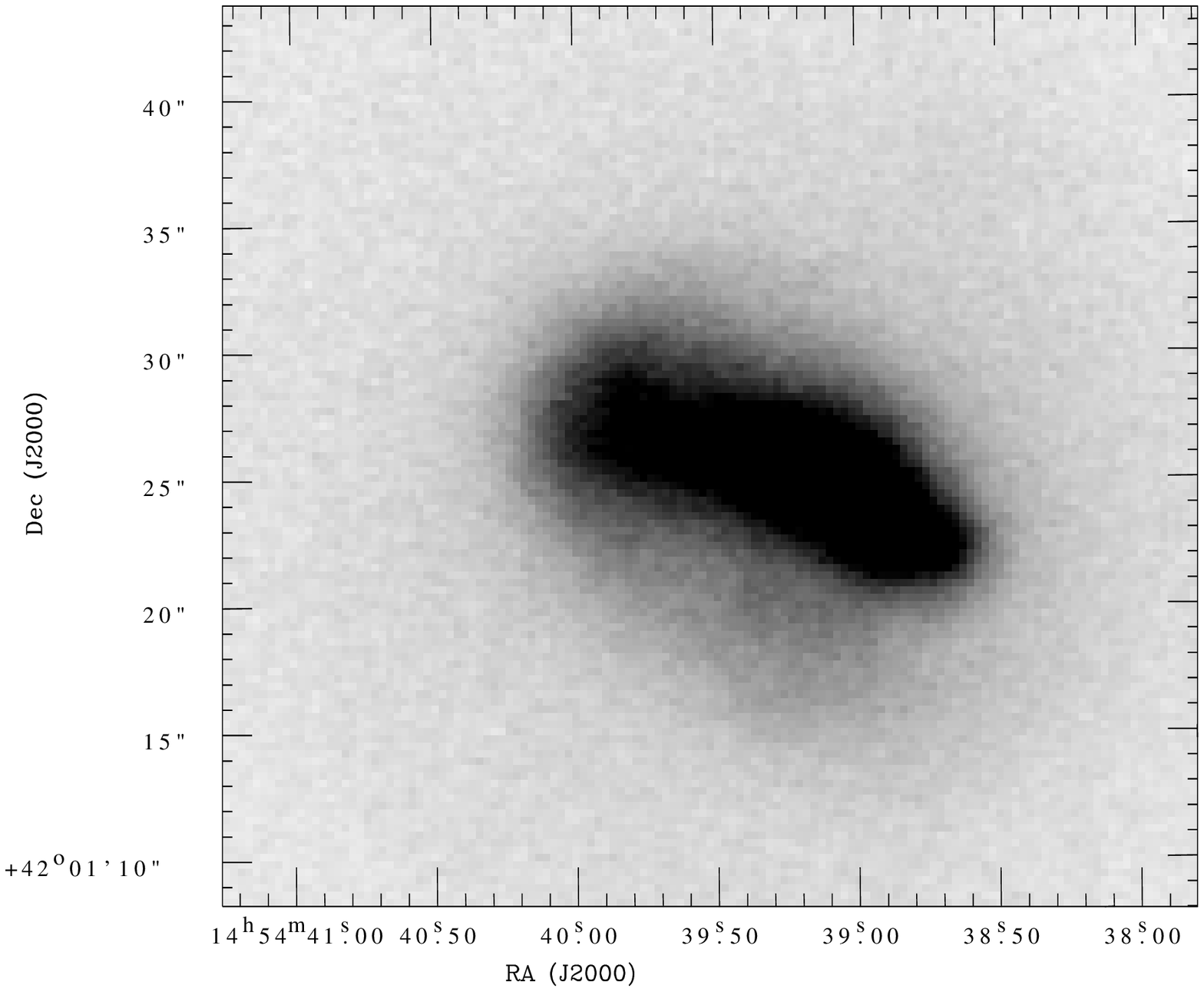}
&
\includegraphics[width=8.1cm]{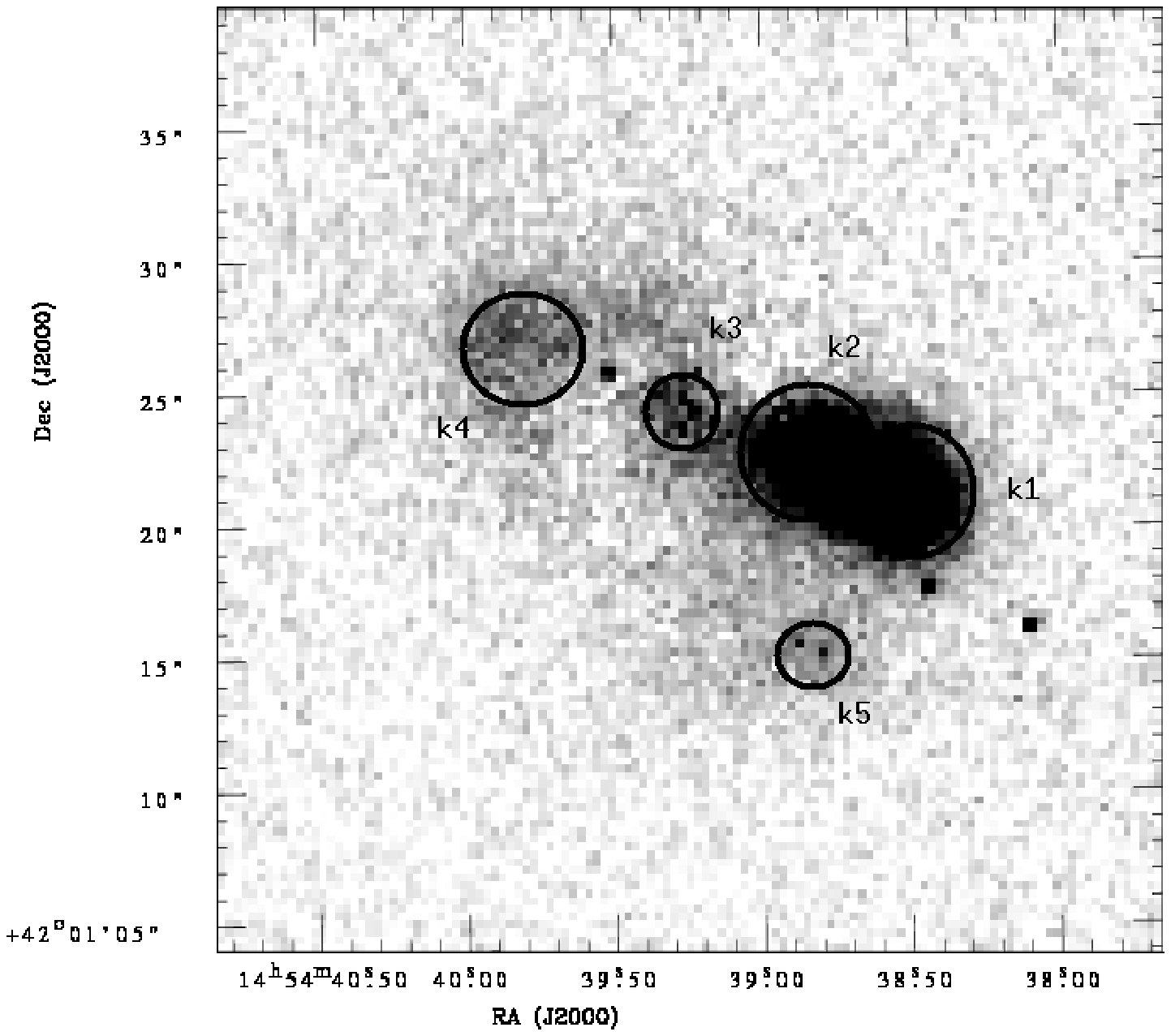} \\
(a) & (b) \\
\multicolumn{2}{c}{\bf I Zw 97} \\
\end{tabular}
\label{f1}
\caption[]{B band (a) and continuum subtracted H$\alpha$ line image (b) of I Zw 97. In both the images north is up and east is left and the field of view is  $\sim1'.0\times1'.0$.}.
\end{minipage}
\end{figure*}

\begin{figure*}
\begin{minipage}{150mm}
\centering
\begin{tabular}{cc}
\includegraphics[width=8cm]{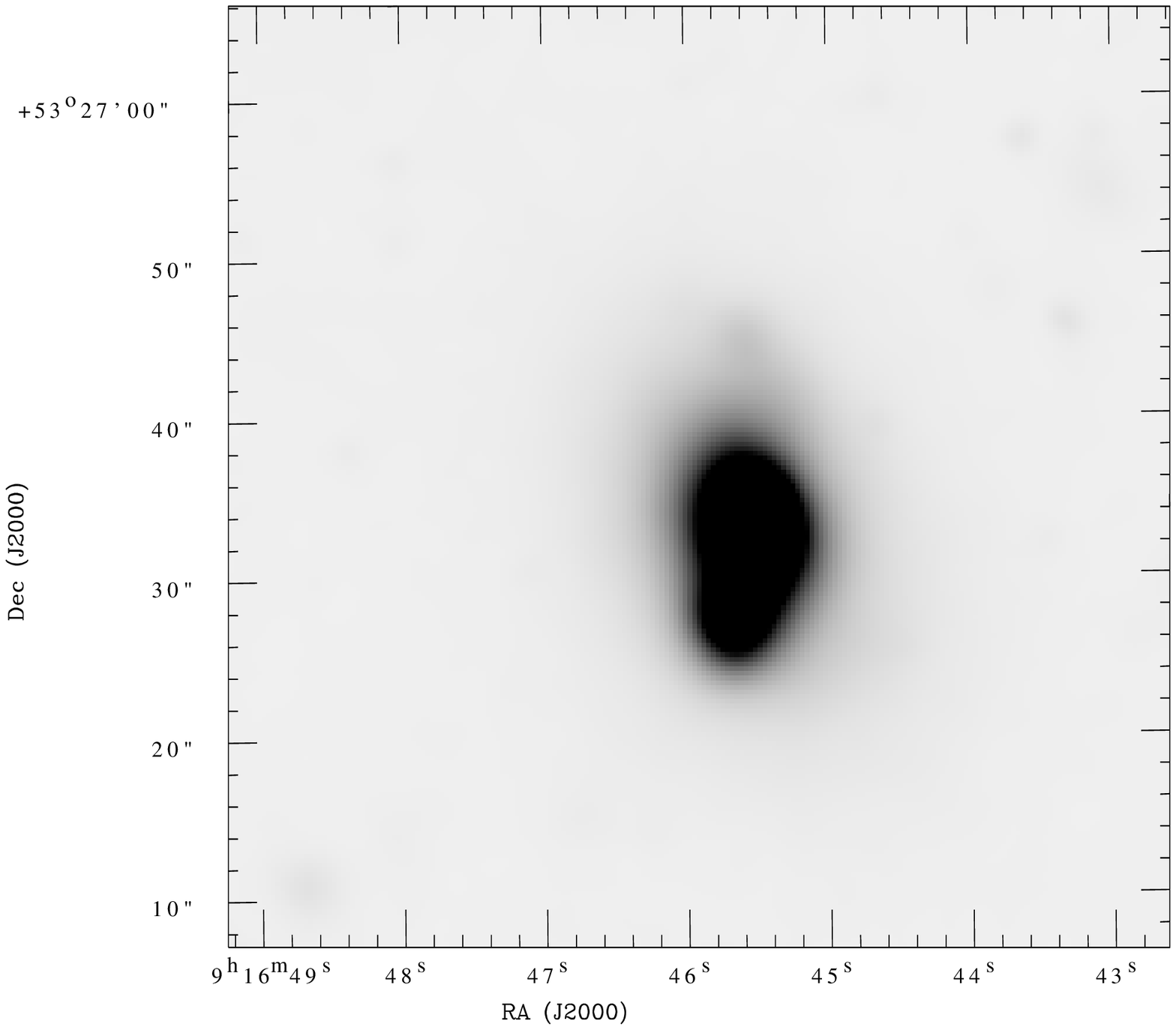}
&
\includegraphics[width=8cm]{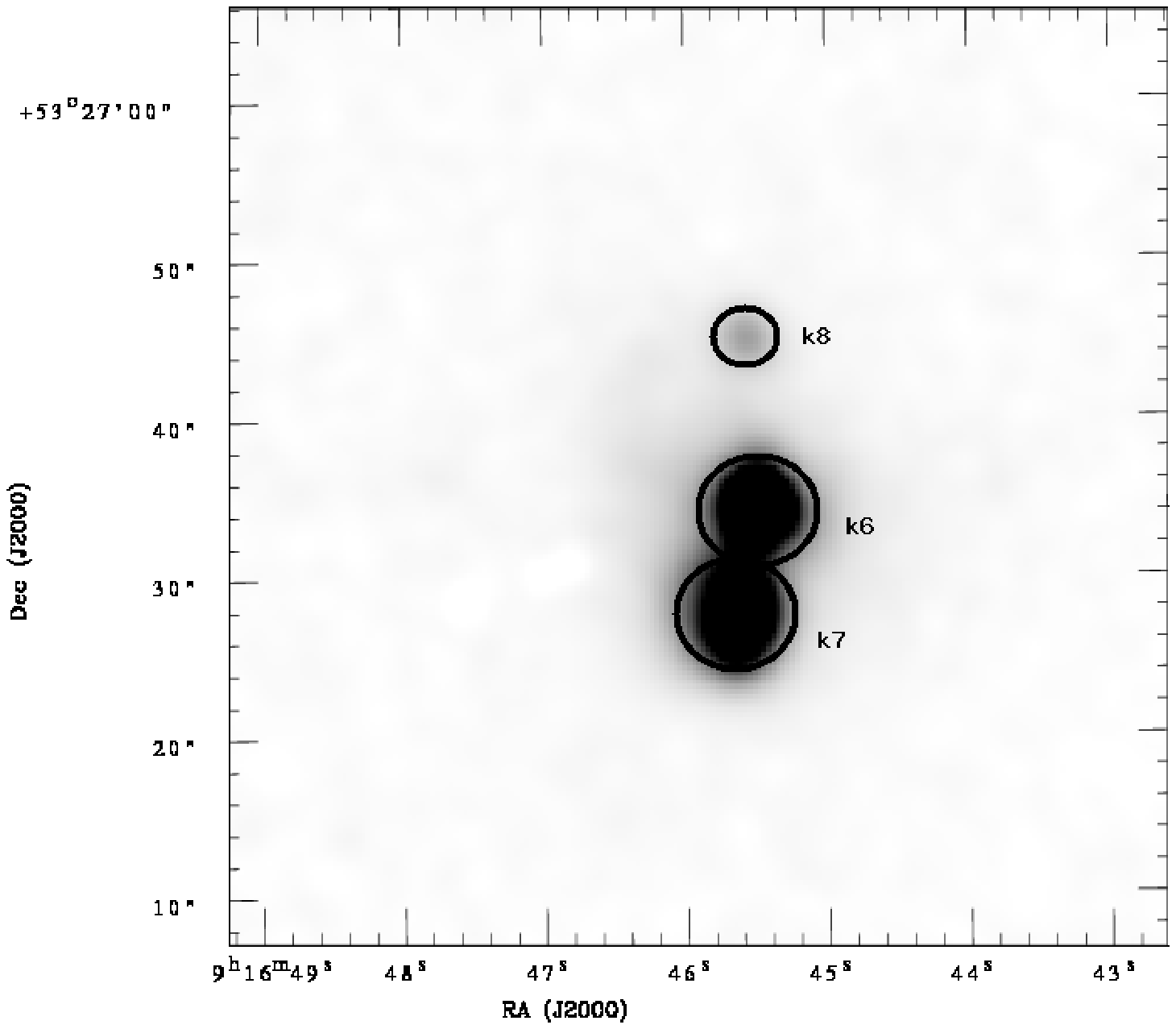} \\
(a) & (b) \\
\multicolumn{2}{c}{\bf Mkn 104} \\
\end{tabular}
\label{f2}
\caption[]{B band (a) and continuum subtracted H$\alpha$ line image (b) of Mkn 104. In both images north is up and east is left and the field of view is $\sim2'.0\times2'.0$.}
\end{minipage}
\end{figure*}

\begin{figure*}
\begin{minipage}{150mm}
\centering
\includegraphics[width=11cm]{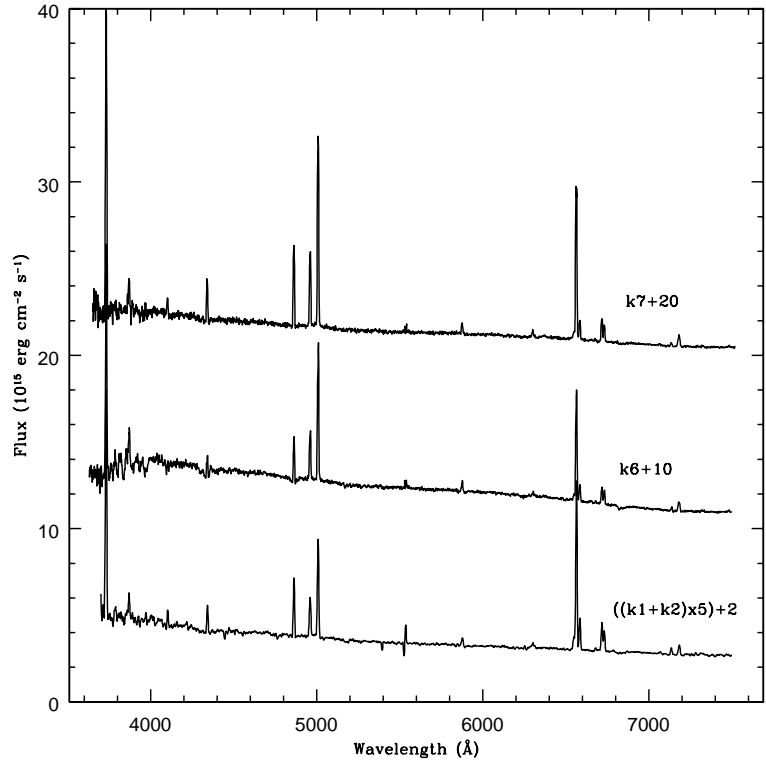}
\label{f3}
\caption[]{Spectra of the bright region k1+k2 (bottom) in  I Zw 97, and knots k6 (middle) and k7 (top) in Mkn 104.}
\end{minipage}\texttt{}
\end{figure*}

\begin{figure*}
\begin{minipage}{150mm}
\centering
\begin{tabular}{cc}
\includegraphics[width=8.1cm]{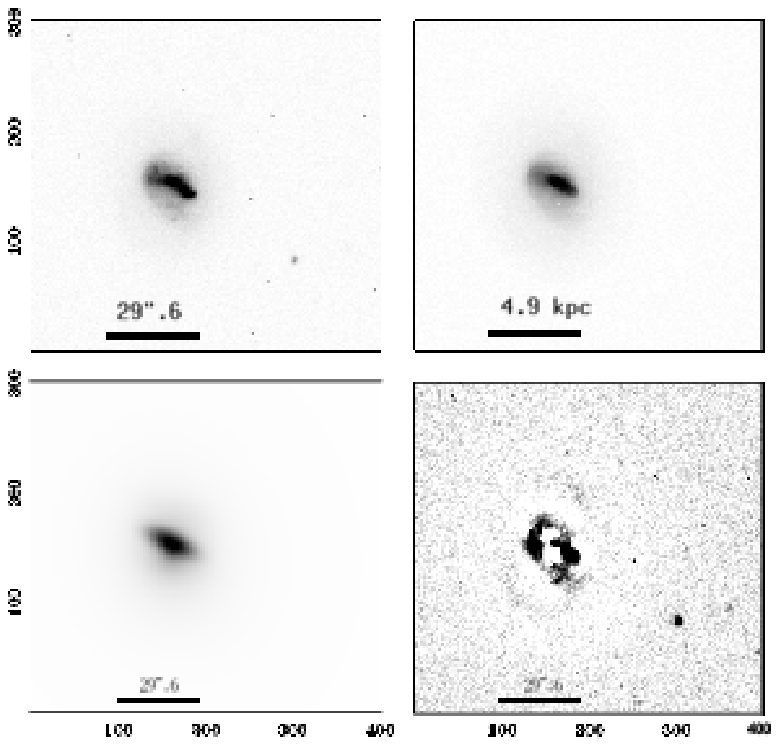} 
&
\includegraphics[width=9.2cm]{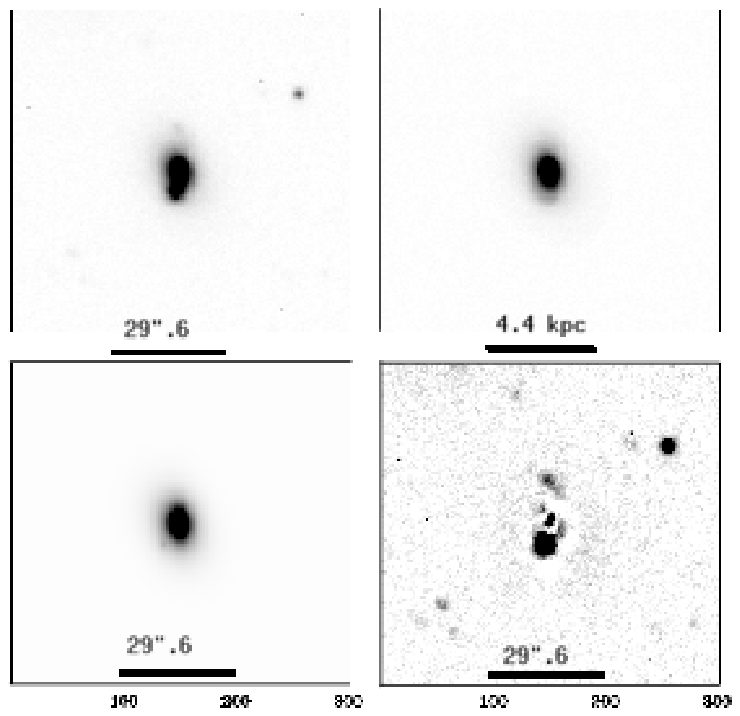} \\
{\bf I Zw 97} & {\bf Mkn 104} \\
\end{tabular}
\label{f5}
\caption[]{For each galaxy, the top left panel shows the $B$ band image, the top right panel shows $B$ band image with H {\sc ii} regions masked, the bottom left shows the model $B$ band image obtained from GALFIT, and the bottom right shows the residual $B$ band image (original $-$ GALFIT model). In all the images north is up and east is left. The field of view of I Zw 97 images are $\sim2'.0\times1'.5$ and of Mkn 104 are $\sim1'.5\times1'.5$.}
\end{minipage}
\end{figure*}

\begin{figure*}
\begin{minipage}{150mm}
\centering
\begin{tabular}{c}
\includegraphics[width=9cm]{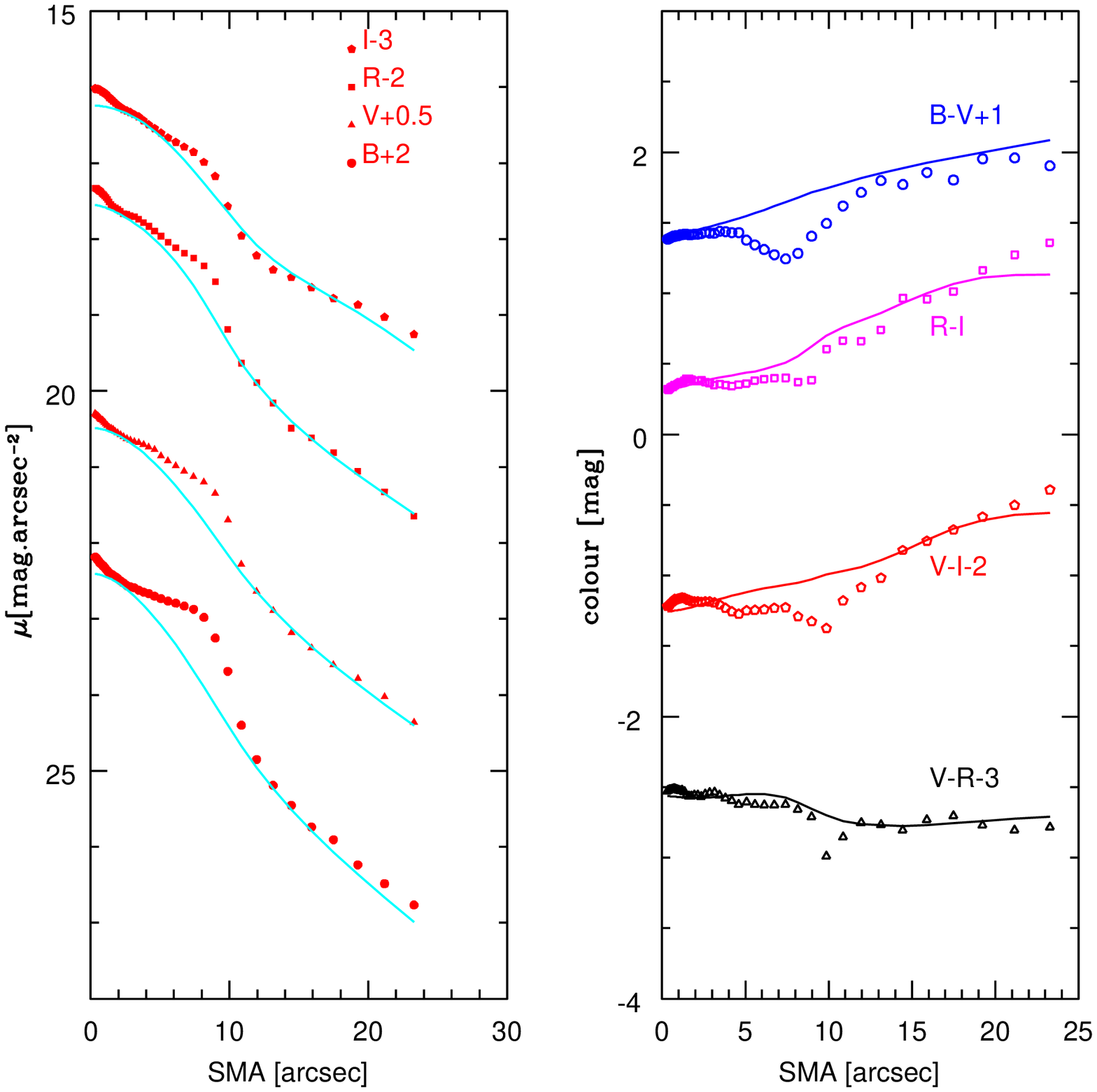} \\
{\bf I Zw 97} \\
\includegraphics[width=9.4cm]{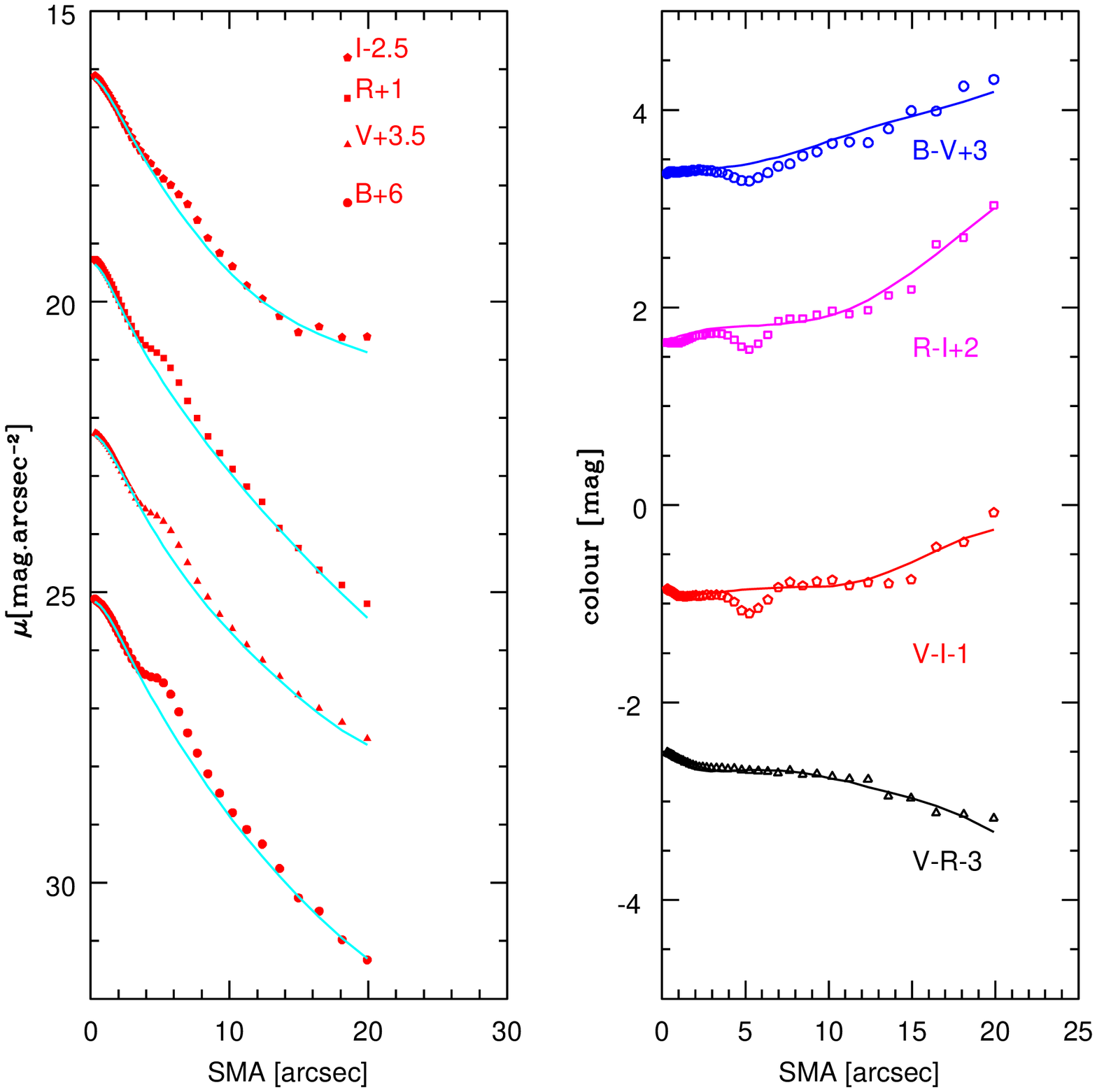} \\
 {\bf Mkn 104} \\
\end{tabular}
\label{f6}
\caption[]{Surface brightness and colour profiles of I Zw 97 and Mkn 104. Points represent the observed surface brightness and colours. Thin lines represent surface brightness and colour profiles of the model galaxy obtained through GALFIT fitting algorithm, where the model galaxy is a smooth version of the original galaxy with prominent star forming regions masked. The surface brightness profile of the model galaxy is reproduced by a combination of two S\'ersic profiles, one representing nearly exponential disk and the other representing a near Gaussian starburst around nucleus. Star forming knots are seen as brightness enhancements and bluer in colour in the observed points.}.
\end{minipage}
\end{figure*}

\begin{figure*}
\begin{minipage}{150mm}
\centering
\begin{tabular}{c}
\includegraphics[width=10cm]{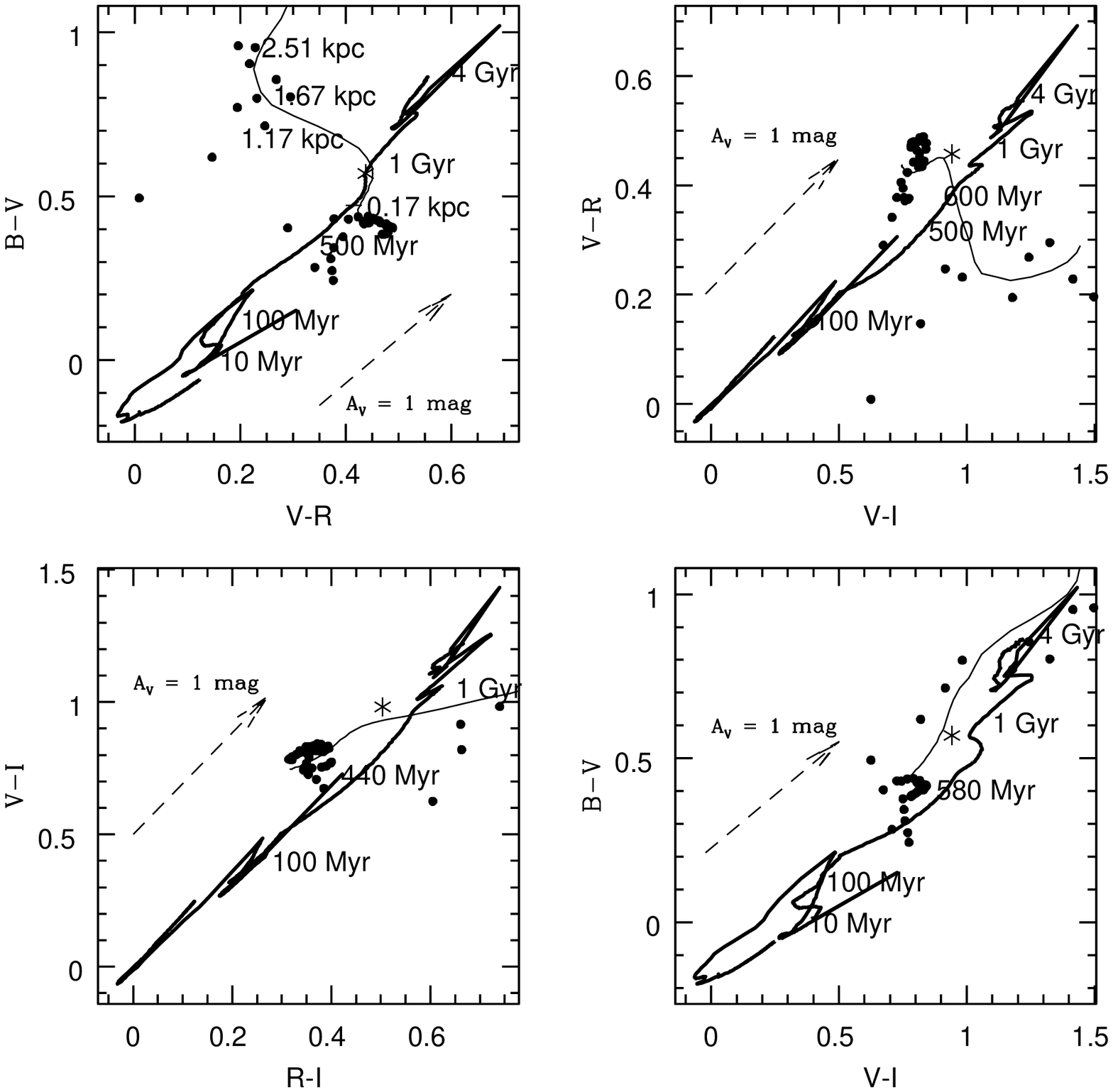} \\
{\bf I Zw 97} \\
\includegraphics[width=10cm]{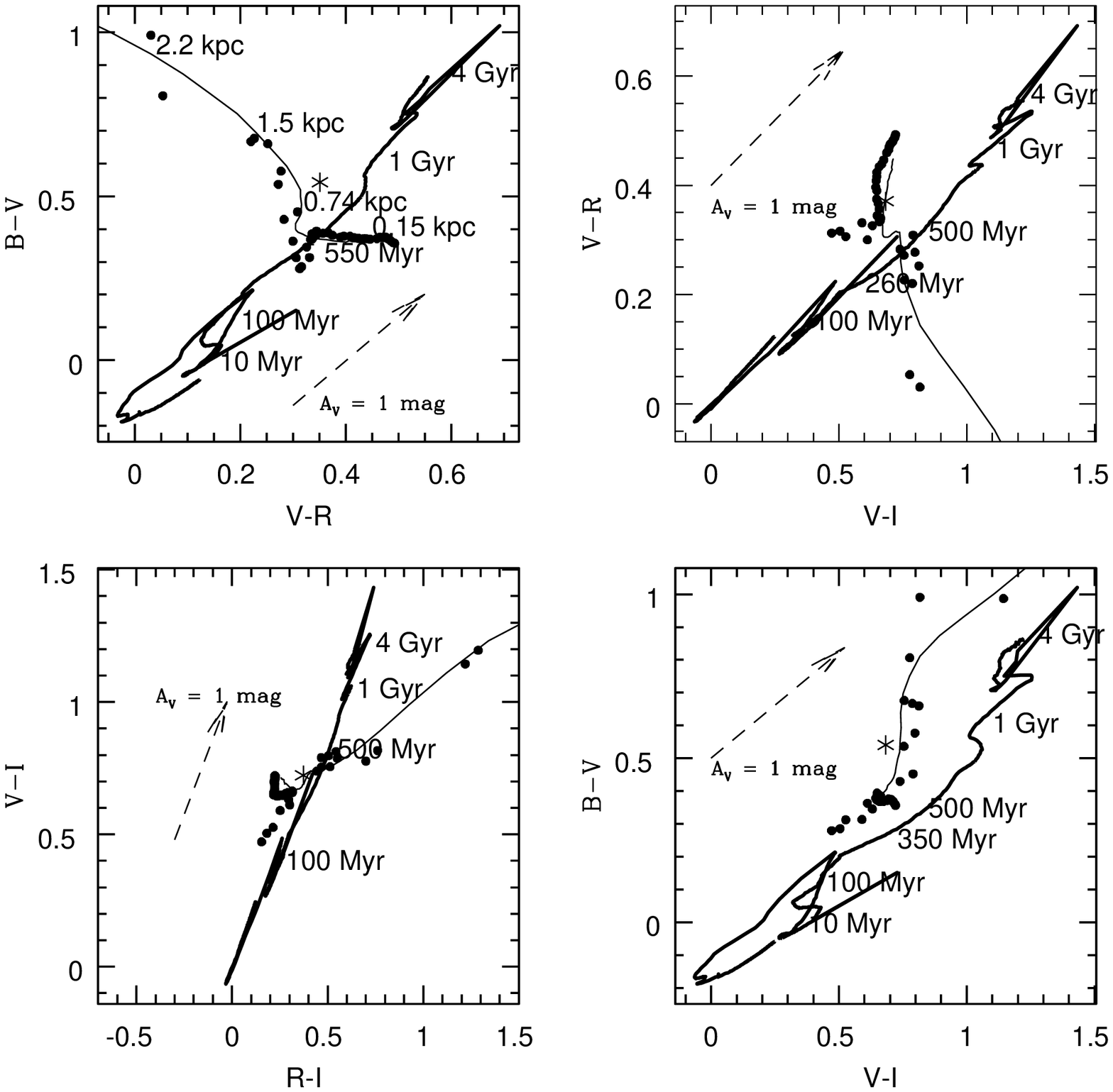} \\
{\bf Mkn 104} \\
\end{tabular}
\label{f7}
\caption[]{Two colour diagram of the colour profiles of I Zw 97 and Mkn 104. Thick line represents the colour-colour plot taken from a Starburst99 model with 0.008 ($Z$) metallicity, Salpeter IMF undergoing an instantaneous burst and evolving from 0.01 Myr - 5 Gyr. Points represent the galaxy colours at different radii from the centre. Thin lines show the colour of the model galaxy obtained through GALFIT fitting algorithm. Probable ages are marked on the Starburst99 evolutionary track. The distance in kpc, from the center of the galaxy are also marked on the thin lines. Mkn 104 has an underlying population of age $\sim550$ Myr and I Zw 97 has an underlying population of age $\sim700$ Myr as seen in the $B-V$ vs $V-R$ diagram. The total integrated broad band colours of the galaxy as obtained from aperture photometry are marked here as $'*'$. Also plotted in these plots are reddening vectors showing A$_V$ = 1 mag extinction.}
\end{minipage}
\end{figure*}

\begin{figure*}
\begin{minipage}{150mm}
\centering
\includegraphics[width=12cm]{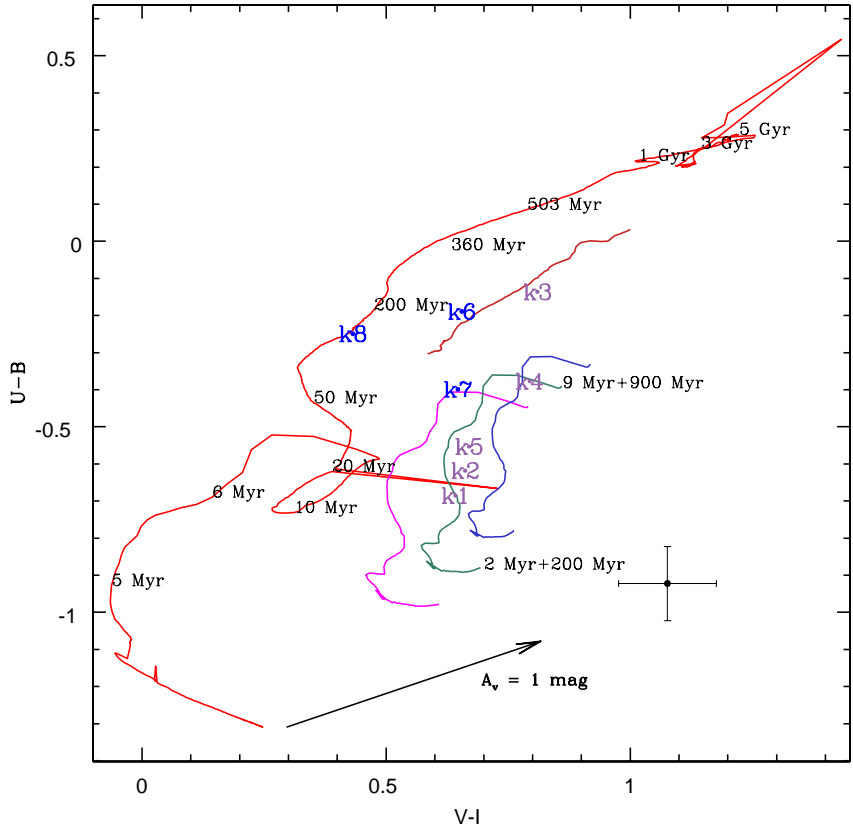}
\label{cc}
\caption[]{ $U-B$ vs $V-I$ colour-colour mixed population model created using a dust free Starburst99 model with a Salpeter IMF and metallicity $Z=0.008$ ($Z_\odot/2.5$). The long red curve represents the evolution of the starburst from 0.01 Myr to 5 Gyr. The short curves refer to loci of points with ages ranging from 2 Myr+200 Myr to 9 Myr+900 Myr mixed with a 4 Gyr population. These are mixed at different fractions as follows. The fraction of young to old (4 Gyr), $f_{2Myr}$ and fraction of 200 Myr to old, $f_{200Myr}$ is 0.007 and 0.05, respectively for the dark blue curve. Fractions $f_{2Myr}$=0.01 and $f_{200Myr}$=0.05 corresponds to the dark green curve, and the magenta curve represents the locus of points having fractions $f_{2Myr}$ and $f_{200Myr}$ as 0.015 and 0.05, respectively. The dark red curve represents the locus of point having fractions $f_{2Myr}$=0.005 and $f_{200Myr}$=0.7. These short curves are best suited to explain the underlying stellar population embedded in the knots k1-k5 of I Zw 97, where marker k3 represents the central region of the galaxy and shows a strong $\sim500$ Myr population.  Knot k7 of Mkn 104 resides in the magenta curve and is explained to contain 4 Gyr population + $\sim500$ Myr + younger $\sim5$Myr population. k6, the central knot of Mkn104, has a similar population to that of knot k3 of I Zw 97. The typical error on these colours (except for knot k8 of Mkn104) is also plotted in the right hand corner of the figure. Since knot k8 of Mkn104 is very faint, the $U-B$ uncertainty is $\sim0.15$ mag and the uncertainty on $V-I$ colour is $\sim0.2$ mag. Also plotted is the reddening vector for an extinction of A$_V$ = 1 magnitude.}
\end{minipage}
\end{figure*}

\begin{figure*}
\begin{minipage}{150mm}
\centering
\includegraphics[width=14cm]{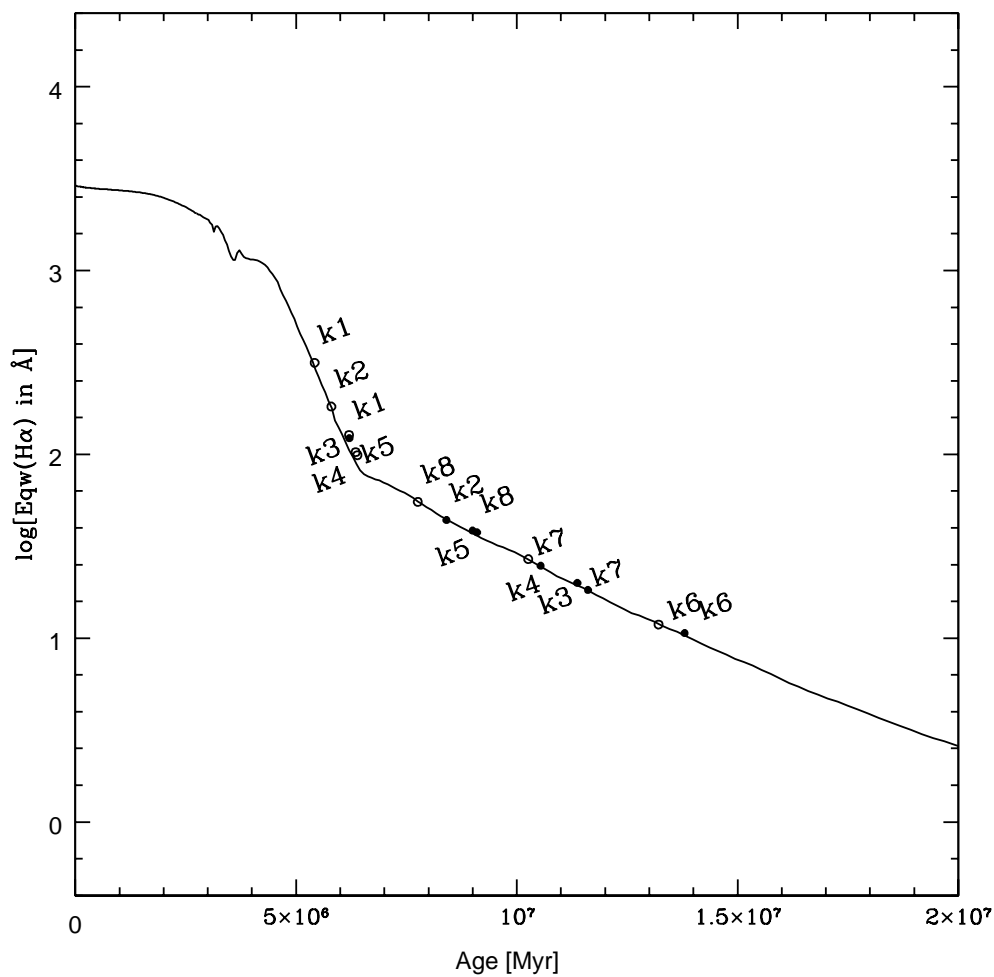}
\label{f8}
\caption[]{Plot of H$\alpha$ equivalent width versus age for the star forming knots. The track is taken from Starburst99 model with 0.008 ($Z$) metallicity and a Salpeter IMF. Closed and open circles represent the corresponding ages of the knots before and after correcting for an underlying old stellar population aged $\sim$ 4 Gyr + 500 Myr in I Zw 97 and in Mkn 104.}
\end{minipage}
\end{figure*}

\begin{table*}
\begin{center}
\caption[]{Log of Imaging and Spectroscopic observations.}
\begin{tabular}{cccc|ccc}
\hline\hline
 &{\bf Date} & {\bf Exp time} & {\bf Seeing} & {\bf Date} & {\bf Exp time} & {\bf Seeing} \\
 &           &   (s)          &              &            &    (s)         &    \\
\hline
\multicolumn{4}{c}{\bf I Zw 97} & \multicolumn{3}{c}{\bf Mkn 104} \\
\hline

$U$ & 15 May 2004 & 1800 & $2\farcs0$ &  07 Jan 2005 & -  & - \\

    & 08 May 2005 & 600 & $2\farcs7$ &  29 Dec 2005 & 840 & $2\farcs8$ \\

$B$ & 15 May 2004 & 1800 & $1\farcs8$ & 07 Jan 2005 & 3180 & $2\farcs2$ \\

    & 08 May 2005 & 360 & $2\farcs3$ & 29 Dec 2005 & 480 & $2\farcs8$ \\

$V$ & 15 May 2004 & 1020 & $1\farcs6$ & 07 Jan 2005 & 900 & $2\farcs0$ \\

    & 08 May 2005 & 240 & $1\farcs9$ & 29 Dec 2005 & 300 & $2\farcs3$ \\

$R$ & 15 May 2004 & 2250 & $1\farcs4$ & 07 Jan 2005 & 1260 & $1\farcs6$ \\

    & 08 May 2005 & 180 & $1\farcs7$ & 29 Dec 2005 & 300 & $2\farcs2$ \\

$I$ & 15 May 2004 &  -  &  - & 07 Jan 2005 & 540 & $1\farcs5$ \\
 
    & 08 May 2005 & 900 & $2\farcs0$ & 29 Dec 2005 & 300 & $2\farcs0$ \\

H$\alpha$ & 08 May 2005 & 1800 & $2\farcs2$ & 9 Jan 2005 & 1800 & $1\farcs5$ \\

167l slit + Grism7 & 25 Mar 2007 & \multicolumn{2}{c}{4200 for knot k1+2*} & 8 Jan 2005 & \multicolumn{2}{c}{2400 for knot k6*} \\
                  &              &                   &                      & 8 Jan 2005 & \multicolumn{2}{c}{2400 for knot k7*}\\
\hline\hline
\multicolumn{7}{c}{* Knots here correspond to the star forming regions marked in Figure 1 \& 2}. \\
\end{tabular}
\label{t1}
\end{center}
\end{table*}

\begin{landscape}
\begin{table*}
\begin{minipage}{180mm}
\caption[]{H$\alpha$ photometric details of the knots. The Eqw({H$\alpha$}$_{cor}$) and Age$_{cor}$ represent H$\alpha$ equivalent width and age after correcting for the underlying old stellar population of 500 Myr + 4 Gyr for I Zw 97 and for Mkn 104.}
\begin{tabular}{p{8mm}|p{13mm}p{14mm}|p{13mm}p{13mm}p{13mm}|p{13mm}p{8mm}p{8mm}p{7mm}|p{10mm}p{13mm}p{13mm}}
\hline\hline
{\bf knot} & {\bf RA} & {\bf Dec} & {\bf F(H$\alpha$)} & {\bf Log(${\rm L}_{\rm H\alpha}$)} & {\bf E(H$\alpha$)} &  {\bf E({H$\alpha$}$_{cor}$)} & {\bf SFR} & {\bf Age} &  {\bf Age$_{cor}$} & {\bf M$_{\rm HII}$} & {\bf log(L$_{IR}$))} & {\bf M$_{old}$} \\

           & hh:mm:ss.s & dd:mm:ss.s & $10^{-14}$ erg/cm$^2$/s & (erg/s) & \AA & \AA & $M_\odot$yr$^{-1}$ & Myr & Myr & $M_\odot$ & $L_\odot$ & $10^9$M$_\odot$ \\
\hline
{ I Zw 97(tot)} & 14:54:39.0 & +42:01:25.0 & 13.78 & 40.29 &  &  & 0.16 &  &  & 2.45e+06 & 9.34 & 2.24 in $H$ band \\
                &            &             &       &        &  &  &       &  &  &          & 9.41  & 2.07 in $K_s$ band \\           
\hline
k1 & 14:54:38.71 & +42:01:22.8 & 3.588 & 39.71 & 122.4 & 315.06 & 0.040 & 6.21 & 5.42 & 6.38e+05  \\
k2 & 14:54:38.86 & +42:01:23.8 & 2.060 & 39.47 & 43.89 & 182.05 & 0.023 & 8.41 & 5.80 & 3.66e+05  \\
k3 & 14:54:39.43 & +42:01:25.8 & 0.571 & 38.91 & 19.97 & 127.40 & 0.006 & 11.37 & 6.20 & 1.01e+05  \\
k4 & 14:54:40.05 & +42:01:29.1 & 0.667 & 38.98 & 24.77 & 102.74 & 0.007 & 10.54 & 6.35 & 1.19e+05  \\
k5 & 14:54:39.00 & +42:01:17.0 & 0.161 & 38.36 & 38.46 & 99.00  & 0.002 & 9.00  & 6.39 & 2.87e+04  \\
\hline
{Mkn 104(tot)} & 09:16:45.5 & +53:26:35.0 & 17.47 & 40.46 &  &  & 0.21 &  &  & 3.62e+06 & 9.15 & 1.45 in $H$ band \\
               &            &             &       &        &  &  &       &  &  &          & 9.25 & 1.41 in $K_s$ band \\
\hline
k6 & 9:16:45.49 & +53:26:37.5 & 1.937 & 39.34 & 10.66 & 11.87 & 0.017 & 13.80 & 13.21 & 2.72e+05  \\
k7 & 9:16:45.59 & +53:26:30.6 & 2.218 & 39.40 & 18.27 & 26.90 & 0.020 & 11.61 & 10.26 & 3.11e+05  \\
k8 & 9:16:45.55 & +53:26:48.1 & 0.169 & 38.28 & 37.59 & 55.13 & 0.002 & 9.10  & 7.76  & 2.36e+04 \\
\hline\hline
\end{tabular}
\label{t2}
\end{minipage}
\end{table*}
\end{landscape}

\begin{table*}
\begin{center}
\caption[]{Photometric details in magnitude units of the knots. Knots k1-k5 corresponds to I Zw 97 and knots k6-k8 belongs to Mkn 104.}
\begin{tabular}{ccccccc}
\hline\hline
{\bf Knot} & $V$ & $U-B$ & $B-V$ & $V-R$ & $V-I$ & $R-I$ \\
\hline
\multicolumn{7}{c}{\bf I Zw 97} \\
\hline
k1 & 17.135 & -0.685 & 0.307 & 0.365 & 0.672 & 0.307 \\ 
k2 & 16.772 & -0.618 & 0.285 & 0.377 & 0.692 & 0.315 \\
k3 & 17.159 & -0.136 & 0.410 & 0.445 & 0.840 & 0.395 \\
k4 & 17.237 & -0.377 & 0.298 & 0.434 & 0.824 & 0.390 \\
k5 & 18.991 & -0.553 & 0.220 & 0.394 & 0.700 & 0.306 \\
\hline
\multicolumn{7}{c}{\bf Mkn 104} \\
\hline
k6 & 15.794 & -0.189 & 0.377 & 0.336 & 0.655 & 0.319 \\
k7 & 15.826 & -0.399 & 0.309 & 0.332 & 0.647 & 0.308 \\
k8 & 18.881 & -0.249 & 0.568 & 0.326 & 0.432 & 0.106 \\
\hline\hline
\end{tabular}
\label{t3}
\end{center}
\end{table*}

\begin{table*}
\begin{center}
\caption[]{Structural parameters of I Zw 97 \& Mkn 104.}
\begin{tabular}{cccccccc}
\hline\hline
{\bf Filter} & {\bf Profile} & {\bf ${\Sigma}_{e}$ \ at \ $r_e$} & {\bf $r_e$} & {\bf $r_e$} & {\bf $n$} & {\bf $q=b/a$} & {\bf P.A.} \\

             &               & mag.arcsec$^{-2}$ &  arcsec     & kpc   &     &         & deg \\
\hline

\multicolumn{8}{c}{\bf I Zw 97}\\
\hline
$B$ & S\'ersic & 20.78 & 5.57 & 0.93 & 0.61 & 0.34 & 68.96 \\
    &          &(0.16) & (0.07)& &(0.01)& (0.00)& (0.12)  \\
    & S\'ersic & 22.64 & 11.33  & 1.89 & 1.01 & 0.92 & 58.71 \\
    &          &(0.16) &(0.12) & &(0.01) &(0.00) &(0.95)  \\
$V$ & S\'ersic & 20.41 & 5.45 & 0.91 & 0.59 & 0.35 & 71.13 \\
    &          &(0.04) &(0.08) & &(0.01) &(0.00) &(0.17)  \\
    & S\'ersic & 22.53 & 12.39 & 2.07 & 0.91 & 0.94 & 64.42 \\
    &          &(0.04) &(0.08) & &(0.01) &(0.00) &(1.17)  \\
$R$ & S\'ersic & 20.11 & 6.03 & 1.08 & 0.41 & 0.31 & 72.20 \\
    &          &(0.04) &(0.13) & &(0.01) &(0.00) &(0.24)  \\ 
    & S\'ersic & 22.02 & 12.76 & 2.13 & 1.21 & 0.92 & 69.38 \\
    &          &(0.04) &(0.22) & &(0.01) &(0.00) &(1.61)  \\
$I$ & S\'ersic & 19.66 & 5.60 & 0.94 & 0.46 & 0.47 & 76.26 \\
    &          &(0.06) &(0.34) & &(0.01) &(0.00) &(1.54)  \\
    & S\'ersic & 21.64 & 15.36 & 2.57 & 0.48 & 0.90 & 73.84 \\
    &          &(0.06) &(0.14) & &(0.02) &(0.01) &(0.88)  \\
\hline
\multicolumn{8}{c}{\bf Mkn 104} \\
\hline
$B$ & S\'ersic & 20.20 & 3.01 & 0.45 & 1.72 & 0.45 & 4.00 \\
    &          &(0.03) &(0.05) & &(0.02) &(0.00) &(0.56)  \\
    & S\'ersic & 21.42 & 6.99 & 1.04 & 0.93 & 0.46 & 23.60 \\
    &          &(0.02) &(0.35) & &(0.02) &(0.00) &(0.13)  \\
$V$ & S\'ersic & 19.77 & 2.90 & 0.43 & 1.30 & 0.53 & 5.10 \\
    &          &(0.02) &(0.15) & &(0.02) &(0.00) &(1.42)  \\
    & S\'ersic & 21.59 & 8.66 & 1.29 & 0.52 & 0.65 & 20.83 \\
    &          &(0.02) &(0.19) & &(0.02) &(0.00) &(0.23)  \\
$R$ & S\'ersic & 19.55 & 3.40 & 0.51 & 1.53 & 0.58 & 4.60 \\
    &          &(0.03) &(0.08) & &(0.02) &(0.00) &(0.81) \\
    & S\'ersic & 21.17 & 8.11 & 1.20 & 0.82 & 0.61 & 25.50 \\
    &          &(0.02) &(0.45) & &(0.04) &(0.00) &(0.44)  \\
$I$ & S\'ersic & 20.01 & 3.90 & 0.58 & 1.47 & 0.58 & 11.34 \\
    &          &(0.04) &(0.24) & &(0.03) &(0.00) &(0.49)  \\
    & S\'ersic & 22.98 & 15.30 & 2.27 & 0.34 & 0.48 & 33.83 \\
    &          &(0.03) &(1.49) & &(0.03) &(0.00) &(0.95)  \\
\hline\hline
\end{tabular}
\label{t4}
\end{center}
\end{table*}

\begin{table*}
\begin{center}
\caption[]{Spectroscopically obtained emission line ratios and abundance estimate of knots k1+k2, k6 and k7.}
\begin{tabular}{cccc}
\hline\hline
 {\bf Ratios} & {\bf k1+k2} & {\bf k6} & {\bf k7} \\
\hline
\ensuremath{\frac{{\rm H}\alpha}{\rm {H\beta}}} & 2.613 & 2.203 & 2.042 \\ \\

\ensuremath{\frac{[{\rm N II}]}{{\rm H}\alpha}} & 0.233 & 0.231 & 0.148 \\ \\

\ensuremath{\frac{ [{\rm S II}]}{{\rm H}\alpha}} & 0.354 & 0.357 & 0.261 \\ \\

\ensuremath{\frac{{\rm [O III]}}{{\rm H}\beta}} & 1.768 & 2.77 & 2.327 \\ \\

\ensuremath{{\rm [S II]}\frac{6717}{6731}} &  1.565 & 1.4 & 1.45 \\ \\

\ensuremath{\frac{{\rm [N II]}}{{\rm [O II]}}} & 0.136 & 0.126 & 0.093 \\ \\

R$_{23}$ & 0.445 & 0.903 & 0.536 \\ \\

log(O/H)+12 & 8.55 & 8.46 & 8.54 \\ \\

\hline\hline
\end{tabular}
\label{t5}
\end{center}
\end{table*}

\end{document}